\documentstyle[aps,twocolumn,prl,epsfig,amssymb,rotating]{revtex}

\def \beq{\begin{equation}}

\def \eeq{\end{equation}}

\def \rpp{R_{\pi \pi}}

\def \beqa{\begin{eqnarray}}
\def \eeqa{\end{eqnarray}}
\begin{document}

\preprint{DPNU-02-30; hep-ph/0209208}
\draft

\title{
\hfill{\normalsize\vbox{\hbox{hep-ph/0209208} \hbox{DPNU-02-30} }}\\
\vspace{-0.5cm}
\bf Determination of weak phases $\phi_2$ and $\phi_3$ \\
from $B\to \pi\pi,K\pi$ in the pQCD method}

\author{\bf Yong-Yeon Keum} 
\address{EKEN Lab. Department of Physics \\
Nagoya University, Nagoya 464-8602 Japan \\
{\it Email: yykeum@eken.phys.nagoya-u.ac.jp}
}

\date{\today}
\maketitle

\begin{abstract}
We look at two methods to determine the weak phases
$\phi_2$ and $\phi_3$ from $B \to \pi\pi$ and $K\pi$ decays within
the perturbative QCD approach. We obtain quite interesting bounds
on $\phi_2$ and $\phi_3$ from experimental measurements in asymmetric 
B-factory: $55^o \leq \phi_2 \leq 100^o$ and $51^o \leq \phi_3 \leq 129^o$.
Specially we predict the possibility of large direct CP violation
effect in $B^0 \to \pi^{+}\pi^{-}$ decay with
$A_{cp}^{dir}(B\to \pi^{+} \pi^{-})=(23\pm7)$ $\%$. 
\end{abstract}


\section{INTRODUCTION}
One of the most exciting aspect of present high energy physics is 
the exploration of CP violation in B-meson decays,
allowing us to overconstrain both sides and three weak phases
$\phi_1(=\beta)$, $\phi_2(=\alpha)$ and $\phi_3(=\gamma)$ of the
unitarity triangle of the Cabibbo-Kobayashi-Maskawa (CKM) matrix
\cite{ckm} and to check the possibility of New Physics.
The ``gold-plated'' mode $B_d \to J/\psi K_s$\cite{sanda},
which allows us to determine $\phi_1$ without any hadron uncertainty,
recently measured by BaBar and Belle collaborations\cite{bfactory}:
$\phi_2=(25.5\pm4.0)^o$.
There are many other interesting channels with which we may achieve this
goal by determining $\phi_2$ and $\phi_3$\cite{gamma}.

In this letter, we focus on the $B \to \pi^{+}\pi^{-}$ and
$K\pi$ processes, providing promising strategies for determining
the weak phases of $\phi_2$ and $\phi_3$, 
by using the perturbative QCD method.

The perturbative QCD method (pQCD) has a predictive power 
demonstrated sucessfully in exclusive 2 body B-meson decays,
specially in charmless B-meson decay processes\cite{pQCD}. 
By introducing parton transverse momenta $k_{\bot}$, 
we can generate naturally the Sudakov suppression effect 
due to the resummation of large double 
logarithms $Exp[-{\alpha_s C_F \over 4 \pi} \ln^2({Q^2\over k_{\bot}^2})]$,
which suppress the long-distance contributions in the small $k_{\bot}$ region
and give a sizable average $<k_{\bot}^2> \sim \bar{\Lambda} M_B$. 
This can resolve the end point singularity problem and 
allow the applicability of pQCD to exclusive B-meson decays. 
We found that almost all of the contributions to the matrix element
come from the integration region where $\alpha_s/\pi < 0.3$ and 
the pertubative treatment can be justified.

In the pQCD approach, we can predict the contribution of non-factorizable
term and annihilation diagram on the same basis as the factorizable one.
A folklore for annihilation contributions is that they are negligible
compared to W-emission diagrams due to helicity suppression. 
However the operators $O_{5,6}$ with helicity structure $(S-P)(S+P)$
are not suppressed and give dominant imaginary values, 
which is the main source of strong phase in the pQCD approach.
Therefore we have a large direct CP violation in 
$B \to \pi^{\pm}\pi^{\mp}, K^{\pm}\pi^{\mp}$,
since large strong phase comes from 
the factorized annihilation diagram, which can distinguish pQCD from
other models\cite{bbns,charm}.

\section{Extraction of $\phi_2(=\alpha)$ from $B \to \pi^{+}\pi^{-}$}
Even though isospin analysis of $B \to \pi\pi$ can provide a clean way
to determine $\phi_2$, it might be difficult in practice because of
the small branching ratio of $B^0 \to \pi^0\pi^0$.
In reality to determine $\phi_2$, we can use the time-dependent rate
of $B^0(t) \to \pi^{+}\pi^{-}$ including sizable penguin
contributions. 
The amplitude can be written by using the c-convention:
\beqa
A(B^0\to \pi^{+}\pi^{-}) &=& V_{ub}^{*}V_{ud} A_u + V_{cb}^{*}V_{cd} A_c
+ V_{tb}^{*}V_{td} A_t,\nonumber \\
&=& V_{ub}^{*}V_{ud}\,\, (A_u-A_t) + V_{cb}^{*}V_{cd} (A_c-A_t),
\nonumber \\
&=& -(|T_c|\,\,e^{i\delta_T} \, e^{i\phi_3} + |P_c|\, e^{i\delta_P})
\eeqa
Pengun term carries a different weak phase than the dominant tree amplitude,
which leads to generalized form of the time-dependent asymmetry:
\beqa
A(t) &\equiv& {\Gamma(\bar{B}^0(t) \to \pi^{+}\pi^{-}) - 
\Gamma(B^0(t) \to \pi^{+}\pi^{-}) \over \Gamma(\bar{B}^0(t) \to \pi^{+}\pi^{-}) + 
\Gamma(B^0(t) \to \pi^{+}\pi^{-})} \\
&=& S_{\pi\pi} \,\,sin(\Delta m t) - C_{\pi\pi}\,\, cos(\Delta m t)
\eeqa
where 
\beq
C_{\pi\pi}={1-|\lambda_{\pi\pi}|^2 \over 1+|\lambda_{\pi\pi}|^2},
\hspace{20mm}
S_{\pi\pi}={2 \,Im(\lambda_{\pi\pi}) \over 1+|\lambda_{\pi\pi}|^2}
\eeq
satisfies the relation of $C_{\pi\pi}^2 + S_{\pi\pi}^2 \leq 1$.
Here 
\beq
\lambda_{\pi\pi} = |\lambda_{\pi\pi}|\, e^{2i(\phi_2 + \Delta\phi_2)}
=e^{2i\phi_2} \left[{1+R_c e^{i\delta} \,e^{i\phi_3} \over 
1+R_c e^{i\delta} \,e^{-i\phi_3} } \right]
\eeq
with $R_c=|P_c/T_c|$ and the strong phase difference
between penguin and tree amplitudes $\delta=\delta_P-\delta_T$.
The time-dependent asymmetry measurement provides two equations for
$C_{\pi\pi}$ and $S_{\pi\pi}$ for three unknown variables 
$R_c,\delta$ and $\phi_2$.

When we define $\rpp=\overline{Br}(B^0 \to \pi^{+}\pi^{-})/
\overline{Br}(B^0\to \pi^{+}\pi^{-})|_{tree}$, 
where $\overline{Br}$ stands for 
a branching ratio averaged over $B^0$ and $\bar{B}^0$, the explicit
expression for $S_{\pi\pi}$ and $C_{\pi\pi}$ are given by:
\beqa
R_{\pi\pi} &=& 1-2\,R_c\, cos\delta \, cos(\phi_1 +\phi_2) + R_c^2,  \\
R_{\pi\pi}S_{\pi\pi} &=& sin2\phi_2 + 2\, R_c \,sin(\phi_1-\phi_2) \,
cos\delta - R_c^2 sin2\phi_1, \\
R_{\pi\pi}C_{\pi\pi} &=& 2\, R_c\, sin(\phi_1+\phi_2)\, sin\delta.
\eeqa
If we know $R_c$ and $\delta$, then $\phi_2$ can be determined 
from the experimental data on $C_{\pi\pi}$ versus $S_{\pi\pi}$. 

Since pQCD provides $R_c=0.23^{+0.07}_{-0.05}$ and $-41^o
<\delta<-32^o$, the allowed range of $\phi_2$ at present stage is
determined as $55^o <\phi_2< 100^o$ as shown in Fig. 1. 
Since we have a relatively large
strong phase in pQCD, 
in contrast to the QCD-factorization ($\delta\sim 0^o$), 
we predict large direct CP violation effect of 
$A_{cp}(B^0 \to \pi^{+}\pi^{-}) = (23\pm7) \%$ which will be tested
by more precise experimental measurement in near future. 
In numerical analysis, since the data by Belle
collaboration\cite{belle} 
is located outside allowed physical regions, 
we considered only the
recent BaBar measurement\cite{babar} with $90\%$ C.L. interval
taking into account the systematic errors:
\begin{itemize}
\item[$\bullet$]
$S_{\pi\pi}= \,\,\,\,\, 0.02\pm0.34\pm0.05$ 
\hspace{8mm} [-0.54,\hspace{5mm} +0.58]
\item[$\bullet$]
$C_{\pi\pi}=-0.30\pm0.25\pm0.04$ 
\hspace{10mm} [-0.72,\hspace{5mm} +0.12].
\end{itemize}
The central point of BaBar data corresponds to $\phi_2 = 78^o$ 
in the pQCD method.

\begin{figure}[thb]
\begin{picture}(200,200)(0,0)
\put(-20,220){
\begin{rotate}{90}
\centerline{\epsfxsize=7.0cm \epsfbox{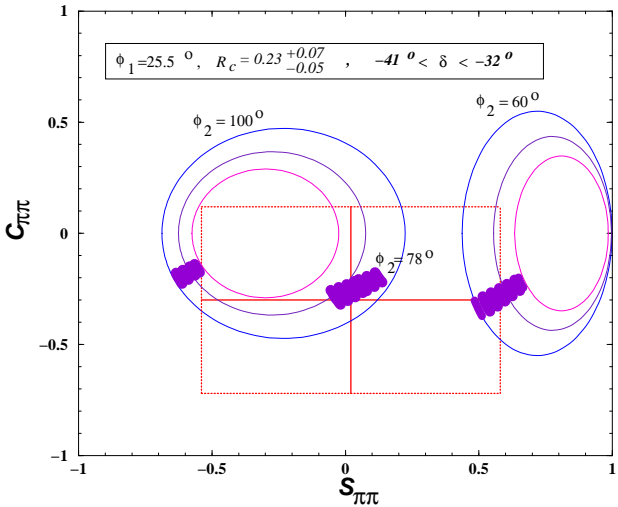}}
\end{rotate}
}
\end{picture} \vspace{5mm}
\caption{Plot of $C_{\pi\pi}$ versus $S_{\pi\pi}$  for various values
of $\phi_2$ with $\phi_1=25.5^o$, $0.18 < R_c < 0.30$ and $-41^o <
\delta < -32^o$ in the pQCD method. Here we consider the allowed experimental
ranges of BaBar measurment within $90\%$ C.L. 
Dark areas is allowed
regions by pQCD for different $\phi_2$ values.}
\label{fig:phi2}
\end{figure}

\begin{figure}[thb]
\centerline{\epsfxsize=8cm \epsfbox{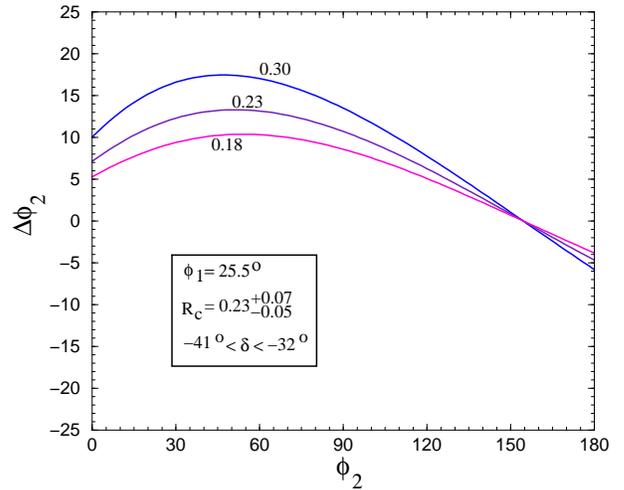}}
\caption{Plot of $\Delta \phi_2$ versus $\phi_2$ with
$\phi_1=25.5^o$, $0.18 < R_c < 0.30$ and $-41^o <
\delta < -32^o$ in the pQCD method.}
\label{fig:delphi2}
\end{figure}

Denoting $\Delta \phi_2$ by the deviation of $\phi_2$ due to the penguin
contribution, derived from Eq.(4), it 
can be determined for known values of $R_c$ and $\delta$ by using
the relation $\phi_3 = 180 -\phi_1 -\phi_2$. 
In Fig.~\ref{fig:delphi2} we show our
pQCD prediction on the relation $\Delta\phi_2$ versus $\phi_2$.
For allowed regions of $\phi_2=(55 \sim 100)^o$, 
we get $\Delta \phi_2 =(8\sim 16)^o$. 
The main uncertainty comes from $|V_{ub}|$ value. 
The non-zero value of $\Delta \phi_2$ demonstrates sizable
penguin contributions in $B^0 \to \pi^{+}\pi^{-}$ decay. 

\section{Extraction of $\phi_3(=\gamma)$ 
from $B^0 \to K^{+}\pi^{-}$ and $B^{+}\to K^0\pi^{+}$ Processes}
By using tree-penguin interference in $B^0\to K^{+}\pi^{-}(\sim
T^{'}+P^{'})$ versus $B^{+}\to K^0\pi^{+}(\sim P^{'})$, CP-averaged
$B\to K\pi$ branching fraction may lead to non-trivial constaints
on the $\phi_3$ angle\cite{fle-man}. In order to determine $\phi_3$,
we need one more useful information 
on CP-violating rate differences\cite{gr-rs02}.
Let's introduce the following observables :
\beqa
R_K &=&{\overline{Br}(B^0\to K^{+}\pi^{-}) \,\, \tau_{+} \over
\overline{Br}(B^+\to K^{0}\pi^{+}) \,\, \tau_{0} }
= 1 -2\,\, r_K \, cos\delta \, \, cos\phi_3 + r_K^2 \nonumber \\
&& \hspace{40mm} \geq sin^2\phi_3     \\
\cr
A_0 &=&{\Gamma(\bar{B}^0 \to K^{-}\pi^{+}) - \Gamma(B^0 \to
K^{+}\pi^{-}) \over \Gamma(B^{-}\to \bar{K}^0\pi^{-}) +
 \Gamma(B^{+}\to \bar{K}^0\pi^{+}) } \nonumber \\
&=& A_{cp}(B^0 \to K^{+}\pi^{-}) \,\, R_K = -2 r_K \, sin\phi_3 \,sin\delta.
\eeqa
where $r_K = |T^{'}/P^{'}|$ is the ratio of tree to penguin amplitudes
in $B \to K\pi$ decay 
and $\delta = \delta_{T'} -\delta_{P'}$ is the strong phase difference
between tree and penguin amplitides.
After eliminate $sin\delta$ in Eq.(8)-(9), we have
\beq
R_K = 1 + r_K^2 \pm \sqrt(4 r_K^2 cos^2\phi_3 -A_0^2 cot^2\phi_3).
\eeq
Here we obtain $r_K = 0.201\pm 0.037$ 
from the pQCD analysis\cite{pQCD} 
and $A_0=-0.110\pm 0.065$ by combining recent BaBar
measurement on CP asymmetry of $B^0\to K^+\pi^-$: 
$A_{cp}(B^0\to K^+\pi^-)=-10.2\pm5.0\pm1.6 \%$ \cite{babar}
with present world averaged value of  $R_K=1.10\pm 0.15$\cite{rk}.

\begin{figure}[thb]
\begin{picture}(200,200)(0,0)
\put(-20,220){
\begin{rotate}{90}
\centerline{\epsfxsize=7.0cm \epsfbox{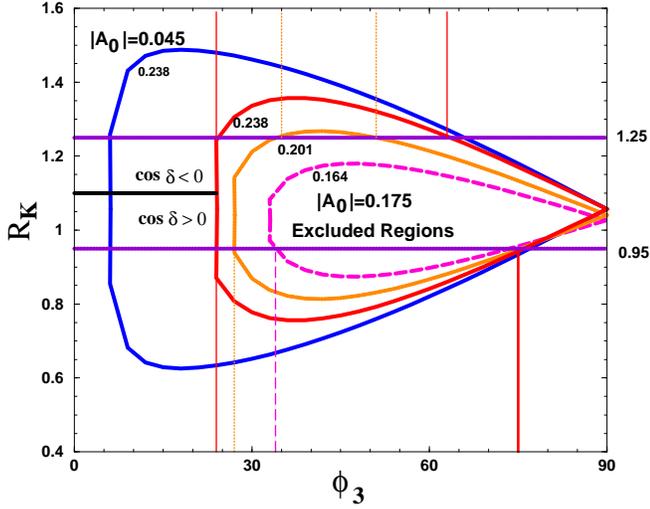}}
\end{rotate}
}
\end{picture} \vspace{5mm}
\caption{Plot of $R_K$ versus $\phi_3$ with $r_K=0.164,0.201$ and $0.238$.}
\label{fig:Rkpi}
\end{figure}
As shown in Fig.~\ref{fig:Rkpi}, we can constrain the allowed range of 
$\phi_3$ within $1\,\sigma$ range of World Averaged $R_K$ as follows:
\begin{itemize}
\item[$\bullet$]For $cos\delta > 0$, $r_K=0.164$: we can exclude
$0^o \leq \phi_3 \leq 6^0$ and $ 24^o \leq \phi_3 \leq 75^0$. 
\item[$\bullet$]For $cos\delta > 0$, $r_K=0.201$: we can exclude
$0^o \leq \phi_3 \leq 6^0$ and $ 27^o \leq \phi_3 \leq 75^0$. 
\item[$\bullet$]For $cos\delta > 0$, $r_K=0.238$: we can exclude
$0^o \leq \phi_3 \leq 6^0$ and $ 34^o \leq \phi_3 \leq 75^0$.
\item[$\bullet$]For $cos\delta < 0$, $r_K=0.164$: we can exclude
$0^o \leq \phi_3 \leq 6^0$. 
\item[$\bullet$]For $cos\delta < 0$, $r_K=0.201$: we can exclude
$0^o \leq \phi_3 \leq 6^0$ and $ 35^o \leq \phi_3 \leq 51^0$. 
\item[$\bullet$]For $cos\delta < 0$, $r_K=0.238$: we can exclude
$0^o \leq \phi_3 \leq 6^0$ and $ 24^o \leq \phi_3 \leq 62^0$.
\end{itemize}

From the table 2 of ref.\cite{keum},
we obtain $\delta_{P'} = 157^o$, $\delta_{T'} = 1.4^o$ and
the negative $cos\delta$: $cos\delta= -0.91$.
The maximum value of the excluded  $\phi_3$ lesser than $90^o$
strongly depends on the value of $|V_{ub}|$.
When we take the central value of $r_K=0.201$,
$\phi_3$ is allowed within the ranges of $51^o \leq \phi_3 \leq
129^o$, because of the symmetric property between $R_K$ vs $cos\phi_3$, 
which is consistent with the result by the model-independent
CKM-fit in the $(\rho,\eta)$ plane.

\section{CONCLUSION}
We discussed two methods to determine the weak phases 
$\phi_2$ and $\phi_3$ within the pQCD
approach through 1) Time-dependent asymmetries in $B^0\to
\pi^{+}\pi^{-}$, 2) $B\to K\pi$ processes via penguin-tree
interference. We can already obtain interesting bounds on $\phi_2$
and $\phi_3$ from present experimental measurements.
Our predictions within the pQCD method is in good agreement with present
experimental measurements in charmless B-decays.
Specially our pQCD method predicted a large direct CP asymmetry
in $B^0 \to \pi^{+}\pi^{-}$ $(23\pm7 \%)$ decay, 
which will be a crucial touch stone
in order to distinguish our approach from others 
in future precise measurements.
More detail works on other methods in $B\to \pi\pi,K\pi$\cite{keum-01}
 and $D^{(*)}\pi$ processes\cite{keum-02}
will appear elsewhere. 
\vskip1cm
\noindent

\acknowledgements
It is a great pleasure to thank 
A.I.~Sanda, E.~Paschos, H.-n.~Li 
and other members of PQCD working group for
fruitful collaborations and joyful discussions.
I would like to thank S.J.~Brodsky, H.Y.~Cheng and M.~Kobayashi for their
hospitality and encouragement. This work was supported in part by
Science Council of R.O.C. under Grant No. NSC-90-2811-M-002 and
in part by Grant-in Aid for Scientific Reserach from Ministry of
Education, Science and Culture of Japan.

\vfil\eject
\end{document}